# Spatially Resolved Sensing in Microfluidics with Multimode Microwave Resonators

Mehmet Kelleci[1,*], Hande Aydoğmuş[1,*], Levent Aslanbaş[1,*], Selçuk Oğuz Erbil[1,*], M. Selim Hanay[1,2,†]

[1] Department of Mechanical Engineering;
[2] National Nanotechnology Research Center (UNAM),
Bilkent University, Ankara, 06800 Turkey

The analogy between mechanical and electromagnetic resonators has been a celebrated paradigm of science and engineering. Exploration of this analogy in recent years has resulted in several exciting research directions, including cavity optomechanics[1], phononic bandgap materials[2] and phononic metamaterials[3-5]. In these examples, progress in electromagnetic research has usually led the way for their mechanical counterparts. Here, we contribute to this analogy from a different perspective by adapting a sensing technique originally developed for mechanical devices to increase the capabilities of sensors based on electromagnetic fields. More specifically, multimode resonance techniques, which enable spatial resolution in inertial mass sensing experiments with nanoelectromechanical systems (NEMS), are tailored for use in microwave resonant sensing, which is commonly employed in microfluidics. We show that the use of higher-order modes of such sensors can provide electrical volume, position and geometric size data. The combination of such spatial features implies the potential for image reconstruction when a large number of modes are used. With the analytical and experimental framework presented here, we can move beyond simple counting and achieve the sizing and imaging of analytes with impedance spectroscopy.

---

[*] These authors have contributed equally to this work.
[†] Corresponding author, e-mail address: selimhanay@bilkent.edu.tr

Microfluidics has emerged in the last two decades as a technology with an immense potential for providing low-cost, mobile and disposable devices for point-of-care screening and diagnostics. To optimize these applications, researchers have invested substantial effort in advancing soft lithography, on-chip fluidic components and microdroplet manipulation technologies. The primary data output from these small devices, however, has been in the form of images obtained using bulky, delicate and expensive optical microscopes. Unfortunately, such reliance on optical microscopy has been one of the major bottlenecks limiting the widespread use of microfluidic devices because of size, portability and cost issues. Therefore, the development of new detection and imaging modalities that are compact and low-cost yet information-rich is critical[6-10] to turn microfluidics technology into an everyday reality.

Electronic detection techniques enable the development of low-cost, integrated sensors[11,12]. In this regard, microwave resonant sensors have many advantages: they can be constructed conveniently, and the resonance frequency of these sensors can be tracked very accurately. At high frequencies, the penetration of electromagnetic fields is not limited by Debye screening[13]. Microwave resonant sensors are sufficiently sensitive to detect individual cells[14], and customized components can decrease costs substantially[15].

Experiments with microwave sensors have thus far focused on the detection of capacitance of analytes near a small electrode. However, a resource that can provide further sensory data remains untapped: *the higher-order modes of a microwave resonant sensor*. The use of higher-order modes (multimode techniques) has substantially advanced the capabilities of nanomechanical sensors. Multimode techniques have been used to locate microparticles[16], weigh single molecules in real-time[17], measure the mass and stiffness of analytes simultaneously[18], obtain spatial information to form an *inertial image*[19], weigh neutral nanoparticles[20] and decouple the effects of multiple particles on the same sensor[21]. These accomplishments have been enabled by the use of multiple modes. In this paper, we explore how similar approaches can be applied to microwave sensors used in microfluidics because the measurement principles and mode shapes of mechanical and electromagnetic resonators are often similar (Figure 1).

Here, we propose to track the resonance frequency of multiple modes of a microwave sensor, record frequency shifts caused by analytes and combine this information in a suitable way to obtain information about the analytes. In principle, each resonant mode provides an independent piece of information. The interaction of an analyte with each resonator mode generates a frequency shift that depends on the spatial overlap between the particle and the electromagnetic mode. The frequency shift data from multiple modes are then processed by an algorithm to obtain the electrical volume of the particle, the centre position, the geometric size of the particle, etc. Unlike capacitance-based sensing, where multiple capacitive electrodes must be fabricated and characterized, the proposed technique uses only a single electrode. As previously demonstrated, the resolution of this technique is not limited by the wavelength of the resonance mode used (because this is not a far-field technique) but rather by the signal-to-noise ratio of the frequency shifts[19]. All of the electrical measurements can be carried out simultaneously using only a single electrode by multiplexing the electronic frequencies.

In dielectric impedance sensing[22], a small particle passing through a channel modulates the effective permittivity of the resonator and induces a shift in the resonance frequency of the mode[23,24]:

$$\frac{\Delta f_n}{f_n} = -\frac{\int_{V_0} \Delta \epsilon(\boldsymbol{r}) E_n^2(\boldsymbol{r}) d^3 \boldsymbol{r}}{\int_{V_0} (\epsilon(\boldsymbol{r}) E_n^2 + \mu(\boldsymbol{r}) H_n^2) d^3 \boldsymbol{r}}$$

where $f_n$ is the original resonance frequency of the mode, $\Delta f_n$ is the change in the resonance frequency (i.e., the signal used for sensing), $\Delta\epsilon$ is the difference between the dielectric constant of the analyte particle and the liquid that it displaces, $\epsilon$ is the dielectric constant of the medium, $\mu$ is the permittivity of the medium, $\boldsymbol{E_n}$ is the electrical field, and $\boldsymbol{H_n}$ is the magnetic field for the $n^{th}$ mode.

We define the LHS of the equation as the fractional frequency shift $\left(\delta f_n \equiv \frac{\Delta f_n}{f_n}\right)$. Furthermore, we note that the denominator in the RHS is the total energy stored in the resonator $E_{res}$ and use the harmonic oscillator property ($< \int \epsilon E_n^2 \, d^3\boldsymbol{r} > = < \int \mu H_n^2 \, d^3\boldsymbol{r} >$) at resonance. The equation can be cast as

$$\delta f_n = -\frac{\int_{V_0} \Delta\epsilon(\boldsymbol{r}) \phi_n^2(\boldsymbol{r}) d^3\boldsymbol{r}}{2V_n} \dots (1)$$

where $V_n$ is the effective electrical volume of the mode: $V_n = \int_{V_0} \epsilon(\boldsymbol{r}) \phi_n^2 d^3\boldsymbol{r}$. In the expression, the overall strength of the electrical field ($\boldsymbol{E_n}$) drops out.[17]

Using this equation as a starting point, one can probe the spatial properties of the permittivity distribution function of a particle ($\Delta\epsilon(\boldsymbol{r})$). To demonstrate how information about $\Delta\epsilon(\boldsymbol{r})$ can be extracted in a model platform, we will consider a microstrip line with a buried microfluidic channel underneath (Figure 1); the analysis can be extended to other types of resonators. We will first describe the point-like particle approximation for the analyte and show how we can detect the position of the particle by two mode measurements. Then, we will discuss how size information can be obtained using four mode measurements.

**Point-Particle Approximation and the Determination of Position**

For a point particle, we can write $\Delta\epsilon(\boldsymbol{r}) = v \, \delta(\boldsymbol{r} - \boldsymbol{r_p})$, where $\boldsymbol{r_p}$ is the position of the particle, $\delta$ is the Dirac delta function and $v$ is the total excess electric volume of the particle. Without loss of generalization, we can consider a one-dimensional microstrip line (Figure 1) as a generic electromagnetic resonator to probe the particle positions along the axial direction. The frequency shifts in the first two modes then read as follows:

$$\delta f_1 = -\frac{v}{2V_1}\phi_1(x)^2$$

$$\delta f_2 = -\frac{v}{2V_2}\phi_2(x)^2$$

For a given platform, the electrical volume of the modes ($V_n$) can be calculated, thereby leaving two unknowns for the problem ($v, x$) and two equations. If the electromagnetic resonator is designed such that $(\phi_1(x)/\phi_2(x))^2$ is an invertible function[17], then these equations can be solved, and the position of the particle can be determined.

As an illustration of the two-mode sensing principle in the point-particle approximation, we study the passage of a droplet along a microfluidic channel. We consider the situation where the microfluidic channel flows directly below the signal path of the microstrip line. A particle placed at different locations will generate different frequency shifts that scale as the square of the local electric field ($E^2(\mathbf{r})$). The electric field has only a z-component directly underneath the microstrip line so that fringing fields can be neglected: $\mathbf{E}(\mathbf{r}) = E(\mathbf{r})\hat{\mathbf{k}}$. Moreover, the electric field will have only a slight variation in the y- and z-directions because the microfluidic channel has a minuscule cross section. In this case, we can express the *n*th mode of the electric field as

$$\mathbf{E}(\mathbf{r}) = A_n\, \phi_n(x)\hat{\mathbf{k}}$$

where $A_n$ is the modal amplitude and $\phi_n(x)$ is the mode shape function for the resonator. For a microstrip line terminated with shorts at both ends, this function can be expressed as

$$\phi_n(x) = \sin(\pi n x)$$

where the spatial coordinate of $x$ is normalised with respect to the length of the microstrip line, L. The frequency shift caused by a particle with an excess dielectric volume ($v$) can be calculated as:

$$\delta f_n = -\frac{v}{2V_n}\sin^2(n\pi x)$$

By using the first two modes and restricting the analysis to the first half of the sensor ($0 < x < 0.5$), we obtain

$$x = \frac{1}{\pi}\arccos\left(\sqrt{\frac{\delta f_2}{4\,\delta f_1}}\right) \quad \ldots (2)$$

After the location is known, then the (excess) electrical volume of the analyte can also be determined by any of the modal equations. One of the hallmarks of two-mode detection is that, when the frequency shift data points are scattered on the two-dimensional plane defined by $\delta f_1$ and $\delta f_2$, each position contour (i.e., the curve on which the value of $x$ is constant) is a straight line passing through the origin. For different position values, the slope of this line changes, as shown before using single-molecule nanomechanical sensors[17].

**Experimental Realization of Position and Electrical Volume Sensing**

To implement the two-mode detection principle in an integrated microfluidic system, we fabricated a microfluidic system embedded in the dielectric region of a microstrip line resonator. The microfluidic system branches into four channels. Each channel crosses the signal path of the resonator at different locations (Figure 2a). To keep the mode shapes as ideal as possible, we fabricated another set of four channels at the other half of the resonator symmetrically, but we did not use them for droplet passage experiments. At the branching points of the channels, droplets selected their paths quasi-randomly so that the droplets passed through all four channels at different times during the tests. Details of the device fabrication are presented in Supplementary Information section S7.

We used water microdroplets generated in an oil flow as analytes. As each analyte passed under the microstrip line, it caused clearly visible frequency shifts on both modes (Figure 2b). The ratio of these frequency shifts signified the location of the droplet. Scatter plots of these frequency shifts clearly show four distinct bands, each corresponding to a different channel (Figure 2c), as predicted by equation (2) and the universal contour plot reported in our previous work with mechanical systems[17]. Independent measurements of the same droplets with a microscope confirmed that the classification based on the frequency shifts agreed with the actual path of the droplets. With a single electronic line, droplet channel assignment was conveniently performed.

The position of a droplet was calculated by the ratio of the frequency shifts. A normalisation factor was applied to the equation to account for the second order effects such as frequency-dependent permittivity and the non-ideal boundary conditions. The resulting histograms agree with the actual locations of the channels in the device within $2\sigma$ error level (Figure 3a and SI section5). As evident from the figure, classification of droplet channels is clearly accomplished and position information is obtained electronically. After the location was known, the electrical volume ($\upsilon$) of the particle was also obtained. As shown in Figure 3b, the histogram for the electrical volume of the droplets shows a sharp peak reflecting the almost monodisperse nature of the droplets produced in the system. This sharp histogram is also significant because the resonant measurements at different locations can be combined to yield the same value for the electrical volume.

We performed FEM simulations and additional experiments using PCBs with drilled holes and a different microfluidic device (Supplementary Information sections 1-3). All these results indicate that the aforementioned formula could be used to correctly calculate the position of the particle. Particle location and electrical volume detection are substantial advancements; the location of a particle throughout a microfluidic channel can be read without a microscope or multiplexed sensor arrays. By using only one electrical conductor and two modes, we could infer the location of the particle in real time. In this manner, the trajectory and speed of a particle can also be determined in real-time along the channel. With one-mode sensing, the trajectory of a particle can be determined only after the particle has passed through the middle point where the maximal value of the frequency shift is used as an indicator. With capacitive sensing, the particle location can be determined when the particle is in proximity with the capacitive electrode. With the two-mode technique, however, the particle location is known at any moment when the channel is placed directly below the microstrip line (e.g., Figure 1). Therefore, accurate position and velocity measurements in applications such as transit time[25] experiments or the real-time feedback control of particle locations[26] can be accomplished without a microscope-based imaging system. More importantly, this work represents the first demonstration of microfluidic microwave sensors with position sensitivity through the device.

**Sizing Analytes and Determining the Relative Permittivity**

When the dimensions of the particle are assumed to have a finite size, then the integral equation presented in equation (1) is more appropriate. The electrical and spatial parameters can be obtained using the corresponding equation for each of the modes tracked. The information from different modes can be used if a suitable superposition is formed. The proof of this approach was provided for mechanical systems in the literature[19] and can be extended to the microwave system presented here, as discussed in the SI.

To demonstrate the calculation of the geometric size (as opposed to the electrical volume, which depends on the electrical permittivity of the analyte), we performed Monte Carlo simulations in Matlab, where the location, size and the permittivity of the particles were randomly assigned. The frequency shifts in the first four modes of a microwave resonator were generated using equation (1) and were then processed via the selection of appropriate values of weights using the approach described in the SI. The electrical volume, location and the variance of the particle were calculated, in this order. The variance values were then converted into sizes under the assumption of a uniform, prismatic shape for the particle. An ensemble of 100 Monte Carlo particles was generated and analysed in this way. The first 15 particles of the ensemble are illustrated in Figure 4. After the geometric size was determined, it was combined with the electrical volume measurements to obtain the effective permittivity of the particle. For the Monte Carlo simulations, the actual and calculated permittivity values for the particles are shown in Figure 4b. Statistics for the performance of the algorithm are presented in the SI. The higher-order modes of the microwave resonator can be incorporated with the same technique[19] to obtain the higher-order moments and, eventually, the image of an analyte.

As evident from simulation results, spatial properties much below the wavelength of modes can be extracted, since the multimode superposition technique is not limited by diffraction effects. In experiments, spatial resolution is usually limited by the frequency noise. An Allan Deviation on the order of $5\times10^{-7}$ is observed during the microdroplet experiments for both modes. By keeping this level of frequency stability, decreasing the electrical volume of the device will improve the achievable spatial resolution by the same amount.

Many challenges should be addressed to advance multimode techniques in electromagnetic resonators. We need to devise strategies to carefully control mode shapes across the device and incorporate higher-order correction terms for changes in mode shapes due to analytes. Moreover, independent calibration is needed for the permittivity of materials at different frequencies so that information from different modes can be combined correctly. Loss tangent of analytes should be considered for more accurate calculations, which changes the frequency of the resonator indirectly by decreasing the Quality Factor. None of these challenges are insurmountable and the study of each one will improve our understanding of microwave sensors.

In conclusion, we have shown how multimode techniques can be transferred from the mechanical domain to the electromagnetic domain by designing a suitable test platform, performing simulations and carrying out two-mode frequency shift experiments with a microfluidic device. Detailed information about the spatial properties of analytes, such as their position, extent, and symmetry, can be obtained using higher-order electromagnetic modes. After these spatial parameters have been obtained, they can be used to reconstruct the image of an analyte passing through a microfluidic channel. With these techniques, size and transit-time measurements can be carried out to gauge the mechanical properties of living cells, e.g., the characterization of single-cell growth rates[27] or the detection of circulating tumour cells (CTCs). Moreover, the permittivity of the particles can also be determined by combining the electrical volume and size measurements, providing means to identify the composition of analytes. By using resonators with mode shapes varying along two- and three-dimensions, images with higher dimensionality can be obtained.

**Acknowledgements**

MSH thanks Barbaros Çetin for formative discussions on microfluidic sensing and for pointing out the potential of this field. MSH thanks John Sader and Michael Roukes for discussions on inertial imaging techniques. We thank Abdullah Atalar and Serkan Kasırga for providing some of the microwave components used in the experiments. We acknowledge the help of Çagatay Karakan, Atakan Arı, Ece Özelçi, Umutcan Çalışkan, Arsalan Nikdoost, Yegan Erdem, Reyhan Tarık, Cenk Yanık and İsmet İnönü Kaya. MS Hanay acknowledges financial support from the European Commission, Marie Curie Program in the form of a Career Integration Grant.

**Author Contributions**

MK, HA, LA and SOE contributed equally to this work.

MSH conceived the idea. All authors took part in designing the experiments. MK performed FEM simulations and initial characterization of microwave sensors. HA, LA and SOE designed and fabricated the microfluidic sensors. MK and SOE performed initial measurements with PCB based microwave sensor. All authors carried out the data acquisition with microfluidic sensors, analysed the data and wrote the manuscript.

**Figures**

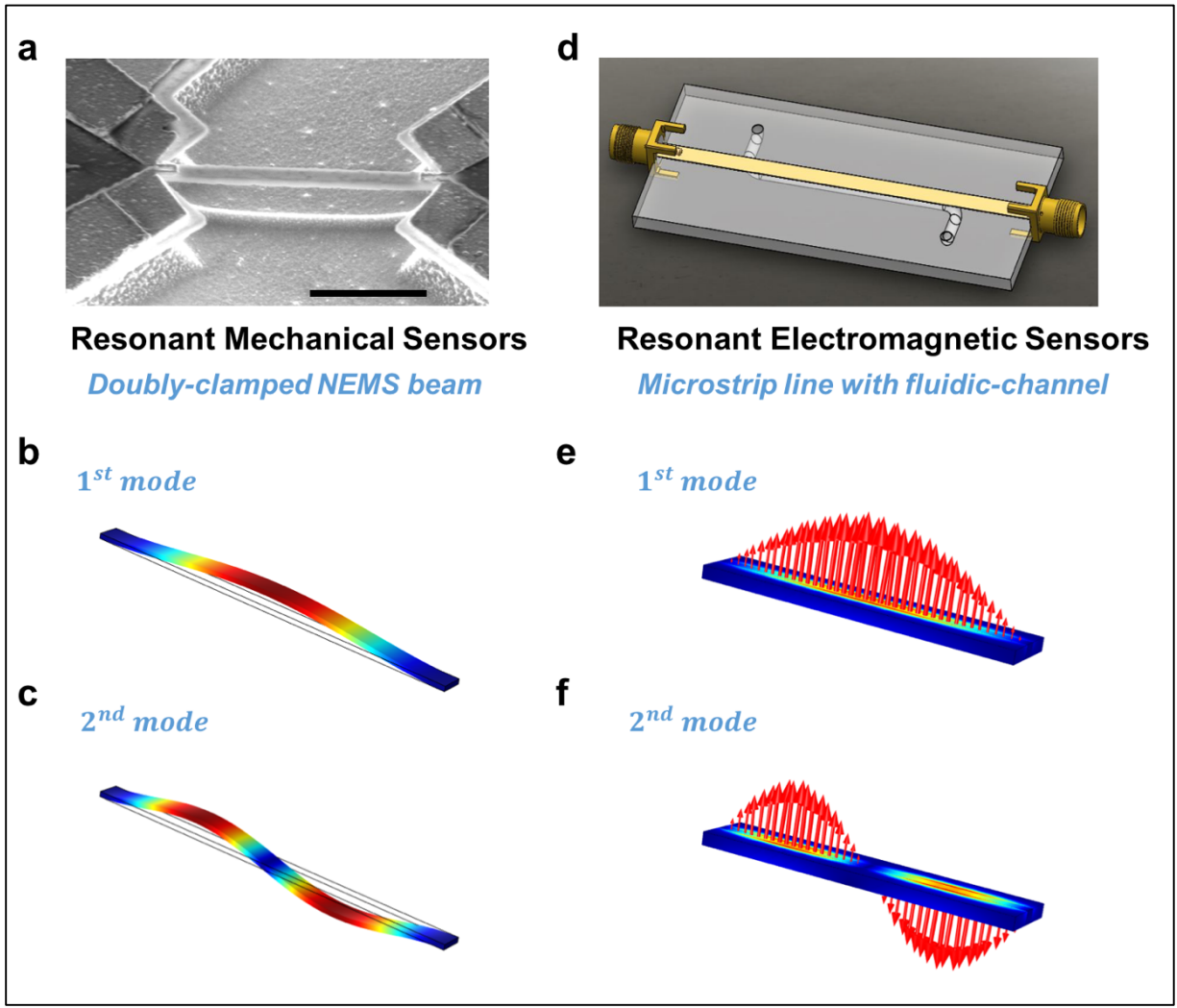

**Figure 1: Mechanical and electromagnetic resonators used as sensors**. In the left panel: **a,** a typical NEMS doubly-clamped beam resonator is shown, along with its **(b)** first, and **(c)** second modes. In the right panel: **d,** an electromagnetic resonator in the form of a microstripline with a buried microfluidic channel is shown. Corresponding electrical fields for the **(e)** first, and **(f)** second modes are also shown on the right panel. The electric field at a given cross-section of the narrow microfluidic channel is almost constant, whereas it changes along the axial direction as shown. Scale bar in (a) is 2 microns.

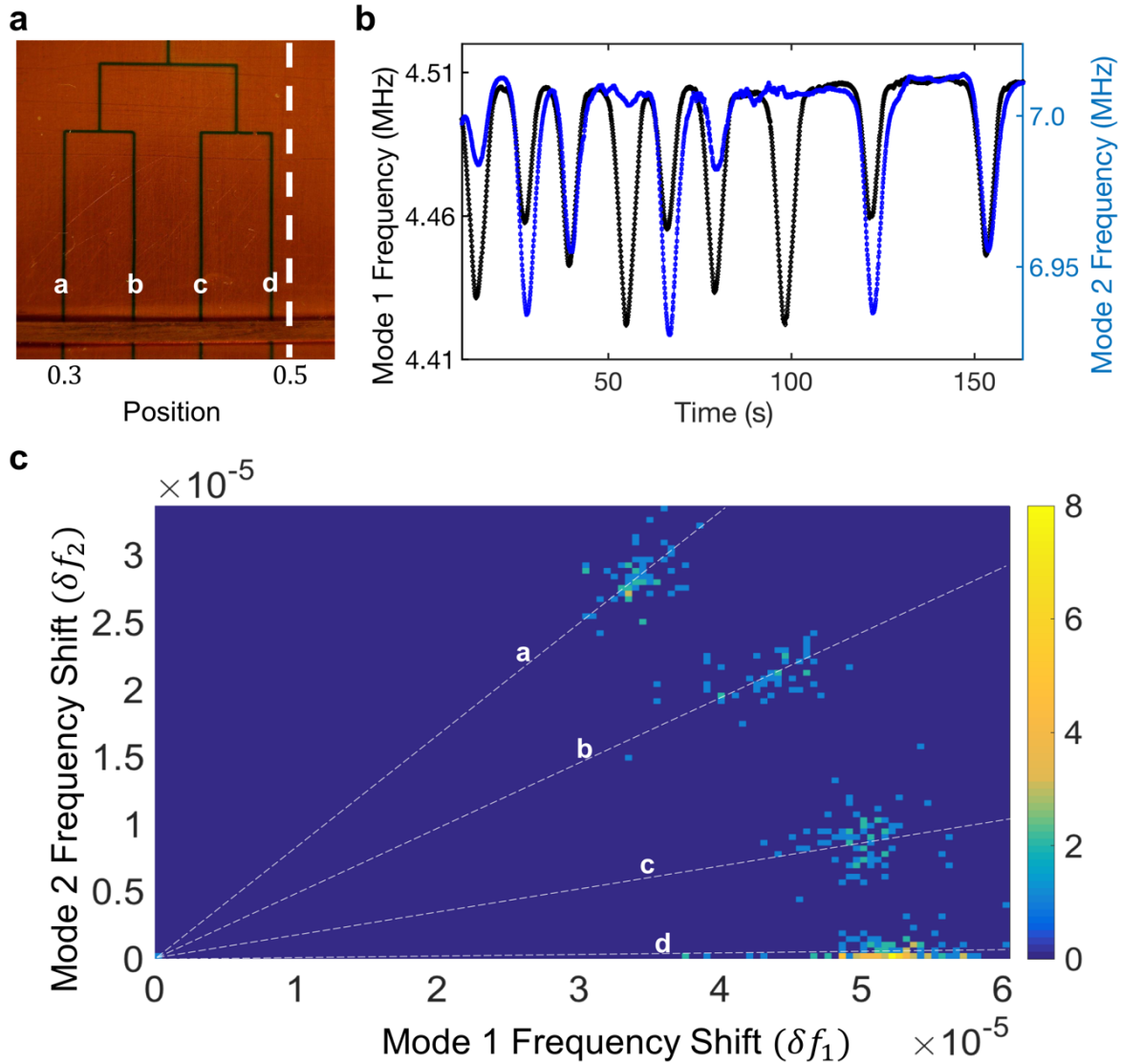

**Figure 2: Two-mode microwave sensor measurements a,** The four microfluidic channels through which the droplets passed are shown. Near the bottom of the image, the copper strip line that formed the signal path of the resonator is visible. The dotted white line indicates the centre of the resonator where the normalised coordinate is at ½, i.e., $x = 0.5$, and the leftmost channel passes near $x = 0.30$. **b,** Two-mode frequency shift data for droplets passing through the system one by one. The ratios of the frequency shifts in the two modes indicate the position of the droplet. For instance, the cases where mode 1 had a large jump mode whereas mode 2 only slightly jumped, (such as when $t \approx 55\ s$ and $t \approx 100\ s$), indicates that the droplet was passing through the node of the second mode i.e., through channel d. Constant offsets of 1.42 GHz from the 1st mode and 2.88 GHz from the 2nd mode were subtracted for clarity. **c,** Scatter plot showing where the two-mode frequency jump events occurred in the $\delta f1 - \delta f2$ plane formed by the normalized frequency shifts for each mode. As droplets passed under the microfluidic channel, they induced frequency shifts in both modes detected with our two-mode PLL system. The white dashed lines were drawn to guide the eye to connect the central region of each cluster to the origin. To elucidate the overlapping events, we presented the data in histogram format.

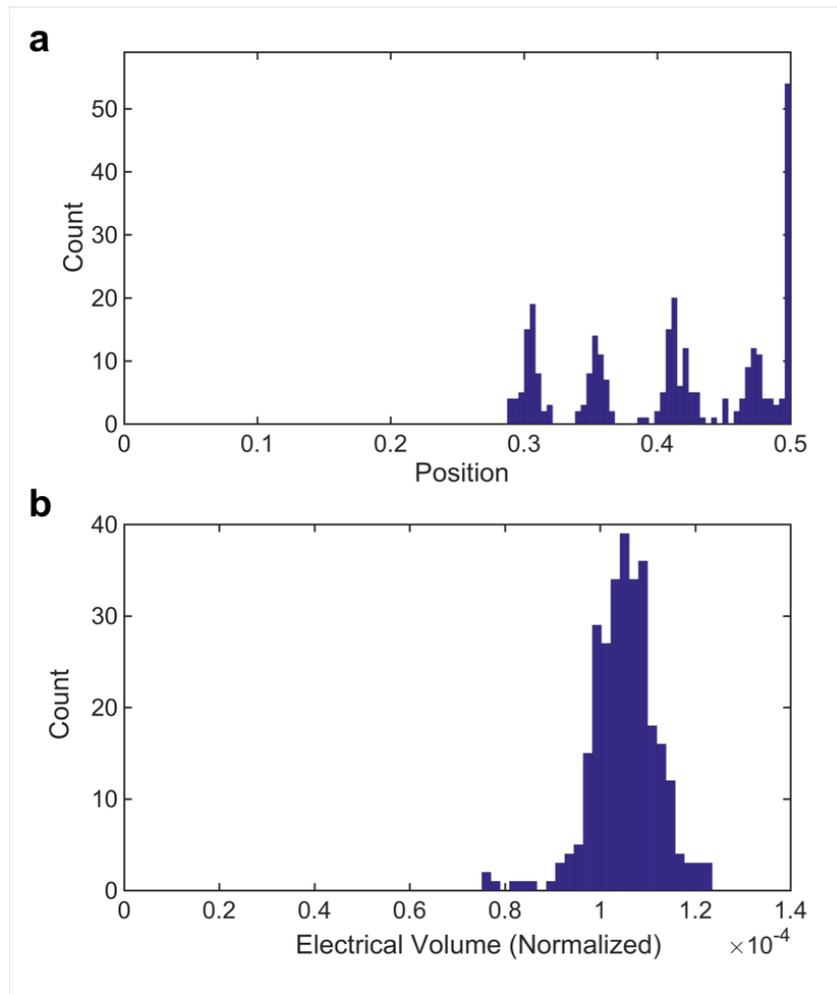

**Figure 3: Measurement Statistics for Microdroplets**. **a,** The histogram of the positions of the droplets. Four different bands are evident, each corresponding to a different channel. The position variables were normalized with respect to the length of the microstrip line so that 0.5 corresponds to the centre of the device. Because the second mode had a node at 0.5 and the fourth channel passed very close to the centre, some of the frequency changes in the second mode were buried in the noise. For these events where the response of the second mode was feeble, the frequency shifts for the second mode were assigned to zero, and the location was accordingly assigned exactly at 0.5. **b,** Measured electrical volume of the droplets normalized to the electrical volume of the device. The sharp peak is an indication of the monodispersity of the generated droplets. The electrical volume was calculated at the frequency of the first mode.

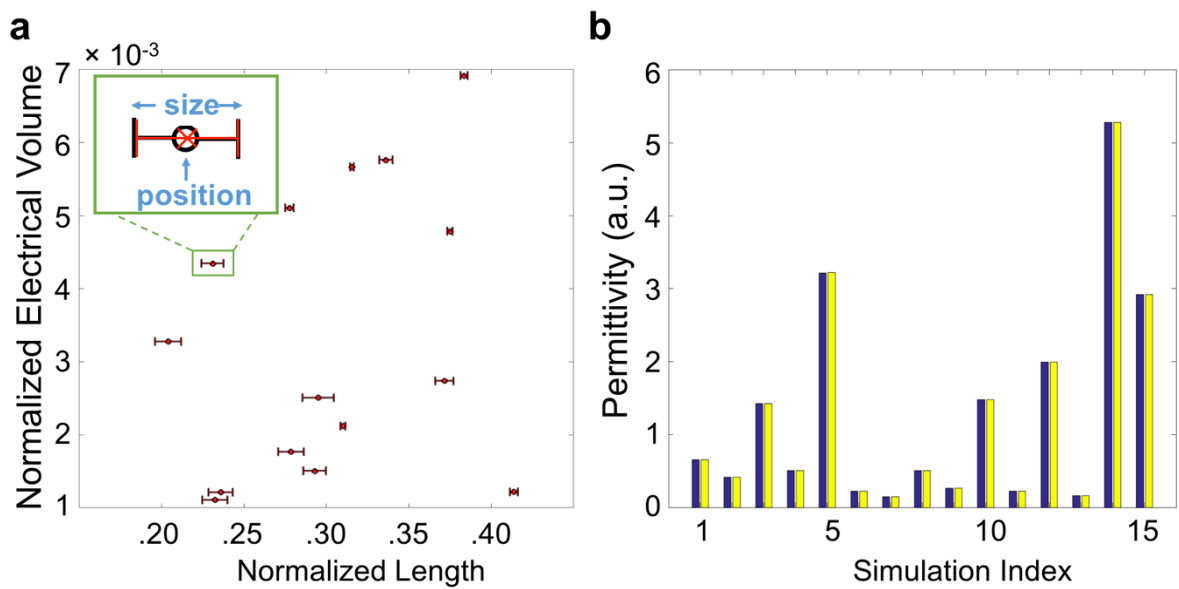

**Figure 4: Size and permittivity determination using four modes**. **a,** Electrical volume, position and size for the simulated (measured) particles are shown in black (red). In most of the cases, the measured values (through the inertial imaging algorithm) overlapped with the actual values. **b,** By dividing the electrical volume of the particle with the geometrical size, the effective permittivity of each particle used in the Monte Carlo simulations was correctly determined. The blue bars show the simulated values, whereas the yellow bars show the calculation results.